# Approaches to Formal Verification of Security Protocols

Suvansh Lal, Mohit Jain, Vikrant Chaplot
Dhirubhai Ambani Institute of Information and Communication Technology
{suvansh_lal, miohit_jain, vikrant_chaplot}@daiict.ac.in

*Abstract*— *In recent times, many protocols have been proposed to provide security for various information and communication systems. Such protocols must be tested for their functional correctness before they are used in practice. Application of formal methods for verification of security protocols would enhance their reliability thereby, increasing the usability of systems that employ them. Thus, formal verification of security protocols has become a key issue in computer and communications security. In this paper we present, analyze and compare some prevalent approaches towards verification of secure systems. We follow the notion of - same goal through different approaches - as we formally analyze the Needham Schroeder Public Key protocol for Lowe's attack using each of our presented approaches.*

## I. INTRODUCTION

Security protocols and algorithms are often used to ensure secure communication in a hostile environment. Today, the need of security and privacy in active information and communication domains such as e-commerce; mobile technologies and the internet has necessitated the use of security protocols. But the underlying security protocols used in these applications are vulnerable to a variety of attacks, such as message replay, parallel session, data interception and/or manipulation, repudiation and impersonation. Therefore before trusting security protocols, it becomes imperative for a systems designer to have some degree of assurance that these protocols fulfill their intended objectives.

Formal verification is the use of mathematical techniques to ensure that a design conforms to some precisely expressed notion of functional correctness. Traditionally, security protocols had been designed and analyzed heuristically. The absence of formal methods for verification could lead to security errors remaining undetected. Formal verification techniques, on the other hand, provide a systematic way of discovering protocol flaws. They can be applied to designs described at many different levels of abstraction, ranging from the gate level, to RTL implementations, and in some cases even to transaction level models described in standardized programming languages [1]. Besides the conformity of correctness, formal methods, when used in the design phases of a system's development often result in more accurate and lower cost systems.

Unfortunately, these protocol verification methodologies are very complex and cannot be easily implemented by protocol engineers [19]. There is also a deep sense of distrust in the academic community about the different verification techniques and their competence in determining prospective design and implementation flaws in complex systems [20, 21]. Therefore through this paper, we intend to present, compare and analyze a few most prevalent approaches towards formal verification of security protocols.

To achieve our objective we introduce four well known verification approaches – the sequential programming approach; the logic programming approach; the strand spaces approach and the belief based approach – falling under the broader domains of **model checking** and **logical inference**. Each of the above mention methodologies are applied to formally verify the Needham Schroeder Public Key protocol [3] for Lowe's attack[4]. In the process of doing so, we compare these approaches for their specification ease, competence in determining complex security flaws; and computational costs. We also make an explicit mention of the advantages and limitations of using these approaches in verifying similar systems.

In section II of the paper we present our literature study by giving a brief mention of what model checking and logical inference approaches are. Section III, introduces the Needham Schroeder Public Key protocol (NSPK) and the classical Lowe's attack on it. In the section IV we elaborate on our presented verification approaches by proving the validity of the Lowe's attack on the NSPK protocol using each approach explicitly. Section V presents a detailed comparison between the approaches explained in section IV. Section VI, gives a brief mention of our intention of future work. Finally, we conclude the paper in section VII, followed by references in section VIII.

## II. LITERATURE SURVEY

Formal verification aims at providing a rigid and thorough means of evaluating the correctness of a security protocol so that even subtle defects can be uncovered. These methods include mathematical analysis dependant on logical analysis or process algebras. Though there are numerous approaches and formal methods that could be employed for verification of a security protocol. These approaches can be broadly classified into two domains namely, model checking and logical inferences.

The first approach is model checking [2], which consists of a systematically exhaustive exploration of the mathematical model. Usually this consists of exploring all states and transitions in the model, by using smart and domain-specific abstraction techniques to consider whole groups of states in a single operation and reduce computing time. Model Checking, one of many formal verification methods, is an attractive and increasingly appealing alternative to simulation and testing to validate and verify systems [2]. Given a system model and desired system properties, the Model Checker explores the full state space of the system model to check whether the given system properties are satisfied by the model. In this paper, we present the logic programming approach [5, 6, 10] under the model checker domain, with *smodels* [25] and *lparse* [12] as a model finder and grounded program generator respectively. Another suggested includes the sequential programming approach [15] with *FDR* [27] as its model checker.

The second approach is logical inference. It consists of using a formal version of mathematical reasoning about the system. This approach is usually only partially automated and is driven by the user's understanding of the system to validate. The suggested approaches under this domain are BAN Logic [22, 23] and the Strand Spaces approach[24].

### III. NEEDHAM SCHRODER PUBLIC KEY PROTOCOL

Proposed by Roger Needham and Michael Schroder, NSPK protocol [3] claims to provide mutual authentication between two agents, along with establishing a session between the communicating parties, in a public key cryptography based system. For the description of the protocol we would assume that A and B are two honest agents and S is a trusted server. The notation used is - $K_{PX}$ and $K_{SX}$ are the public and private keys of agent X.

**Protocol Run:**

1) A → S: A, B (A requests B's public key from S)

2) S → A: {$K_{PB}$, B}$K_{SS}$ (S responds. B's identity is send along with $K_{PB}$ for confirmation)

3) A → B: {Na, A}$K_{PB}$ (A sends a fresh nonce Na to B)

4) B → S: B, A (B requests S for A's public key)

5) S → B: {$K_{PA}$, A}$K_{SS}$ (S sends the public key of A to B)

6) B → A: {Na, Nb}$K_{PA}$ (B generates a fresh nonce Nb and sends it back to A, along with A's nonce Na)

7) A → {Nb}$K_{PB}$ (A confirms Nb to B)

At the end of the protocol, A and B know each other's identities, and know both Na and Nb. These nonces are not known to eavesdroppers.

**Attack on the NSPK protocol:** Gavin Lowe proposed an attack on NSPK using CSP modeling technique and FDR model checking tool [4]. Lowe claimed that the protocol is vulnerable to Man in the Middle Attack, wherein an adversary I, who is responding to a protocol run initiated by A, can falsely authenticate itself to an agent B as A, by replaying A's message to B. Thus, B is fooled to belief that a session is established between A and B. In the explanation of the attack and its analysis, we will ignore the messages transmitted to and from a trusted server S (message 1, 2, 4 and 5) which remain unchanged in an attack run.

**Lowe's Attack on NSPK:**

1.1) A → I: {Na, A}$K_{PI}$ (A sends a fresh nonce Na to I)

2.1) I(A) → B: {Na, A}$K_{PB}$ (In a parallel run of the protocol, I masquerading as A, relays the message received from A after encrypting it under B's public key.)

2.2) B → I(A): {Na, Nb}$K_{PA}$ (B responds to I's message)

1.2) I → A: {Na, Nb}$K_{PA}$ (I relays B's message to A)

1.3) A → I: {Nb}$K_{PI}$ (A returns Nb to complete protocol run with I)

2.3) I(A) → B: {Nb}$K_{PB}$ (I masquerade A and forwards Nb encrypted under B's public key)

In the next section, different verification approaches are applied to formally analyze NSPK protocol for a security violation that corresponds to Lowe's attack and produce results similar to those stated above.

### IV. APPROACHES TO FORMAL VERIFICATION

#### A. The Strand Spaces Approach

Strand Spaces proposed by Fábrega et al in [24] is a mathematical technique for formal verification of security protocols. A *strand* represents the chronological sequence of the messages transferred during a protocol run. These messages can be sent or received by either legitimate parties or the adversary. The collection of strands of all the parties participating in the protocol run is known as a Strand Space.

This technique provides a distinguished approach for protocol verification with intelligent and reliable proofs even without *automated support*. It works with the explicit model of possible behavior of system penetrator and also provides clear semantics about data items like nonce and session keys. It also provides proofs of notions of correctness of both secrecy and authentication. It also provides the detailed insight into why certain assumptions are required to prove the correctness.

TABLE 1
TERMINOLOGY USED

| | |
|---|---|
| A | Message Space. It has two disjoint subsets:<br>• T: set of atomic text messages<br>• K: set of cryptographic keys |
| <σ,a> | Signed Term: + represent sent message, - represents received message. |
| (tr, Σ) | Trace mapping of a participant. It represents the set of messages sent or received by a participant. |
| ⊑ | Subterm relation. e.g. m ⊑ {m}K but K ⋢ {m}K iff K ⊑ m |

| n1 ⇒ n2 | Node n1 is the immediate causal predecessor of n2 |
| --- | --- |
| n1 → n2 | There is a causal link between the nodes n1 and n2 |

**Proof of Needham – Schroeder Public Key Protocol**

PROTOCOL AIM:
This protocol intends that after the successful run of the protocol, the communicating parties share access to $N_a$ and $N_b$ and no other party should have access to these values.

DEFINITION: Let $\sum$ be the NSPK strand space. It is the union of following strands:
- Initiator Strand: The Initiator Strand has the *trace* Init [A, B, $N_a$, $N_b$].
- Responder Strand: The Responder Strand is complementary to the Initiator strand and has the *trace* Resp [A, B, $N_a$, $N_b$].
- Penetrator Strand P.

ASSUMPTIONS
- Each participant has the knowledge of other participant's public key.
- Each participant has different public key. According to this condition if $K_A = K_B \Rightarrow A = B$ and vice versa.
- Nonce values $\{N_a, N_b\} \neq \{A, B\}$. This implies that name of any participant is not used as nonce value.

PROOF
Figure 1 represents the ideal NSPK strand space. Here 'i' represents the initiator strand while 's' represents the responder strand.

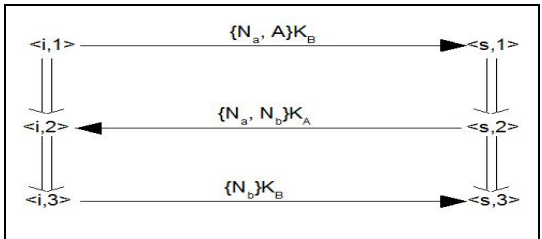

**Figure 1**

PROPOSITION 1: Responder's Guarantee
*Given*
- $\sum$ is the NSPK strand space and C is a bundle that has the responder strand s with trace Resp[A, B, $N_a$, $N_b$];
- $K_A^{-1} \notin K_P$ where $K_P$ represents the set of keys know to penetrator P;
- $N_a \neq N_b$ and $N_b$ originates uniquely in $\sum$.

*Then there exists an initiator strand 'i' in bundle C with trace Init [A, B, $N_a$, $N_b$].*

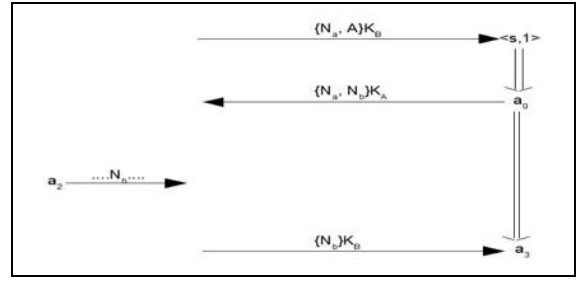

**Figure 2**

Figure 2 represents the legitimate responder's strand.
To prove the proposition we will take help of certain lemmas. We rename node <s, 2> where responder sends message $\{N_a, N_b\}_{KA}$ as $a_0$ and its term as $t_0$. We also refer node <s, 3> where responder receive the message $\{N_b\}_{KB}$ as $a_3$ and its term as $t_3$.

**LEMMA 1.1**: $N_b$ *originates at* $t_0$.

*Proof:* By observing Figure 2 we find that $N_b \sqsubset a_0$ which is a positive node. Now the only node preceding $a_0$ on responder strand is <s, 1>. We just have to check that $N_b \not\sqsubset$ term<s, 1>, i.e. $N_b \neq N_a$ and $N_b \neq A$. Both of these follow from our hypothesis. Hence we can say that $N_b$ originates at $a_0$.

Next lemma checks that whether the step 3 of the protocol is taken by the legitimate party or the penetrator.

**LEMMA 1.2**: *Given set* $S = \{ n \in C : N_b \sqsubset term(n) \wedge t_0 \not\sqsubset term(n)\}$. *Set S has a minimal node* $t_2$. *The node* $t_2$ *is positive and regular.*

*Proof:* It can be seen that $a_3 \in S$. Hence S is non empty and contains a minimal node $a_2$. Since $a_2$ is minimal node its sign is positive. We will now check that whether $a_2$ lies on a penetrator strand or not. We will examine the possible types of penetrator traces.

<+t> : If trace(p) is of this form than $N_b = t$. This implies that $N_b$ originates on this strand which contradicts the result of Lemma 1.1

<-g>: This trace lacks any positive node and hence cannot be minimal node.

<-g, +g, +g>: In this case positive nodes do not have minimal occurrence.

<-g, -h, +gh>: In this case positive nodes do not have minimal occurrence.

<+$K_0$>: Here $K_0 \in K_P$. But since from our assumption, text and keys belong to disjoint sets this cannot be possible.

<- $K_0^{-1}$, -$\{h\}_{K0}$, +h>: If the positive node is minimal then $t_0 \not\sqsubset$ but $t_0 \sqsubset \{h\}_{K0}$. Hence $K_0 = K_A$. This contradicts our assumption that $K_A^{-1} \notin K_P$

Therefore $a_2$ does not lie on a penetrator strand but it must belong to a regular strand.

**LEMMA 1.3**: *A node* $a_1$ *precedes* $a_2$ *on the same regular strand and term($a_1$) = $\{N_a, N_b\}_{KA}$.*

$N_b$ originates uniquely in $\sum$. Also $a_2 \neq a_0$ which follows from above proved lemma. Therefore there should be some node $a_1$

preceding $a_2$ such that $N_b \sqsubset$ term $(a_1)$. Since node $a_2$ is minimal term $(n_1) = \{N_a, N_b\}_{KA}$.

**LEMMA 1.4**: *The regular strand i containing $a_1$ and $a_2$ is the initiator strand and this strand has trace Init [A, B, $N_a$, $N_b$]*.
With the weaker information proved in above lemmas we cannot conclude that i has a trace of the form Init [A, B, $N_a$, $N_b$]. This is because the responder's identity is not determined by the term $\{N_a, N_b\}_{KA}$, which is what this agreement protocol is all about. We can only conclude that strand t belonging to user A has trace Init [A, X, $N_a$, $N_b$] where X can be some party with which A is communicating. This results corresponds to the Lowe's attack suggested in section III.

### B. THE LOGIC PROGRAMMING APPROACH

In this section we use ALSP (*Action Language for Security Protocols*[5, 6] as an efficient specification language to formally analyze the NSPK protocol[3]. ALSP is an executable language for representing protocols and checking for security violations they are vulnerable to [6, 10]. Specification of a protocol in ALSP requires inculcation of concepts of robotic planning[13]. Security protocols are reframed as planning problems, where agents exchange messages and are subject to attacks by intruders. The specification of a protocol in ALSP is viewed as a plan to achieve a goal and the attacks become plans, which achieve goals that correspond to security violations.

ALSP is based on LPSM(*Logic Programming with Stable Model Semantics*) [7, 9]. Logic Programming[8] enables one with declarative ease to specify the actions of the different agents in a protocol. This includes both the operational behavior of a protocol, along with the possible security attacks of an intruder. All stable models[11] for the solution set of logic programs in ALSP are *minimal* and *grounded* in nature[9]. Minimalism allows one to determine exactly what happened when a protocol specification was executed. It ensures that all unwanted models are not a part of our solution set. Groundedness, on the other hand ensures that everything present in the solution set has a justification behind its presence[5].

A logic program is written as a set of *Horn clauses* known as *rules*. We formalize and frame all actions in a protocol specification as logic programming rules[6]. A rule comprises of a *head* and a *body*, separated by a [:-]. The left hand side *head* literals hold true if all the literals on the right hand side *body* are true. A syntactically correct example of a rule in logic programming would be:
```
q:- p, s.
```
Let *P* be the logic program with *S* being the solution set for *P*. Then the above rule could be read as, if the literals *s* and *p* belong to the solution set *S* then *q* must also belong to the solution set *S*. Here the rule is a constraint on the solution set.

To develop the protocol specification for NSPK, initially a general description of the background and action theories pertaining to the protocol is specified in ALSP. This is followed by defining, the choice rules representing the correct execution of the protocol and the actual specification of the protocol dependant itself. Last but not the least; we define a rule corresponding to the security property we want to check. Then, the above specifications are merged and a maximum execution time for the protocol run is determined. Setting bounds on basic objects like nonces, devices, random values, etc we use *lparse*[12] as a suitable front end to the *smodels*[25] system to generate a grounded logic program from our specification. Finally this grounded logic program is executed in *smodels* to find stable models corresponding to violations.

**Development of Protocol Specification for Needham – Schroeder Public Key Protocol:**

We start specifying the protocol by considering basic sort predicates to characterize the basic components of NSPK. We state clearly the background theory (initial state of the protocol) which contains *rules* describing, how a message is composed and decrypted by the agents. It also includes the properties of keys shared and how information is attained by agents participating in a protocol. A few basic sort predicates used in our protocol specification are, `nonce(N)`, `agent(A)` `time(T)`. The names of these predicates are intuitive and represent properties and functions of these predicates. A special sort predicate `msg(M)` is also defined, which means that *M* is a valid message that may appear in a protocol run. Then, we specify a few basic constructors that symbolize cryptographic operations, concatenations and hashing of messages as required by the protocol. Table 2, represents a few classical constructs used in the protocol specification.

TABLE 2
CONSTRUCTS USED IN NSPK SPECIFICATION

| `encrypt(K, M)` | Denotes an encryption of the message M using symmetric key K. |
|---|---|
| `concat(A, B)` | Denotes concatenation of messages *A* and *B* (A\|\|B) |

We also specify predicates that define the properties of messages and keys that are used in the protocol. In addition to this, definition of the ability of agents to construct, send, receive and understand these messages is also an imminent part of our protocol specification. As suggested in [6], the predicate names in most part of our ALSP specification for NSPK are fairly intuitive and represent the action or property after which they are named. Table 3, gives a brief mention of these predicates.

TABLE 3
PREDICATE NAMES AND THEIR FUNCTIONS

| `part(M,M1)` | Denotes *M* as a submessage of *M*1 |
|---|---|
| `verifier(V,A,B)` | Denotes *V* as a verifier shared between agents *A* and *B* |
| `knows(A,M,T)` | Denotes agent *A* knows message *M* at time *T* |
| `synth(A,M,T)` | Denotes agent *A* synthesizes message *M* at time *T* |

| `says(A,B,M,T)` | Denotes agent *A*'s attempt of sending the message *M* to agent *B* |
|---|---|
| `gets(A,M,T)` | Denotes agent *A*'s receipt of message *M* at time *T* |

Having known all the constructs and predicates required for our specification of NSPK. We initiate the specification of the background theory, beginning with the messages that can be used in the protocol. In ALSP it is deemed sufficient to specify the valid messages, along with all their sub messages for a protocol run [13,14]. This is an important step which ensures that the ALSP specification is admissible [6, 10]. NSPK being a three step protocol, we can easily distinguish the three messages along with their sub messages, transmitted at different stages of its execution. To enhance the readability of our specification, we have avoided using the sort predicates in the body clause of our rules. We have also used the more prevalent notation(Table 1) for its description henceforth. For example, xor(M1,M2) has been simply written as *M*1 ⊕ *M*2 and concat(M1,M2) as *M*1||*M*2.

The protocol specification should also represent the ability of agents to modify and manipulate messages.

```
msg(encrypt(K, N, A)):-
          nonce(N), agent(A), publicKey(K, A).
msg(encrypt(K, N)):-
          nonce(N), publicKey(K, A).
msg(N||A)):-
          nonce(N), agent(A).
```

To do this message parts are inductively defined based on the protocol constructors. Most of this specification is independent of the protocol itself and represents message part defining rules, incorporated from [5].

```
part(M,M) :-msg(M).
part(M,M1||M2):- msg(M), msg(M1),msg(M2),part(M,M1).
part(M, M1||M2):- msg(M),msg(M1),msg(M2),part(M,M2).
```

Modeling of knowledge is also an important aspect of the protocol specification in ALSP. Intuitively, the *knows* predicate is used for the purpose. This includes modeling the abilities of agents to acquire information from messages they have either received or transmitted. We also define that if an agent possesses the knowledge of a message *M*2 then, he/she would also possess the knowledge of a message *M*1 which is a sub message of *M*2. This enables the agent to extract useful message parts from concatenated or e*xored* messages.

```
knows(A, M, T):- said(A, B, M, T)
knows(A, M, T):- got(A, M, T)
knows(A, M, T):- knows(A, M1, T), part(M, M1)
knows(A, M, T):- knows(A, M1||M2, T), part(M, M1)
```

Similarly we specify the ability of an agent to synthesize a message in a protocol run. The *rules* defining the synthesis of messages ensure that an agent can construct a message if and only if it can construct and thereby knows, all the subparts of that message.

```
synth(A, M, T):-knows(A, M, T)
synth(A, prf(M1, M2), T):-knows(A, M1, T),
                          knows(A, M1, T)
```

Knowledge modeling, description of messages and specification of the ability of agents to synthesize valid messages in a protocol run concludes our background theory for NSPK. Next, comes the specification of the action theory for the protocol. Most of the specification in this part is protocol independent and we refer to [6] for a detailed description.

```
got(B, M, T+1):-gets(B, M, T)
said(A, B, M, T+1):-says(A, B, M, T)
got(B, M, T+1):-got(B, M, T)
said(A, B, M, T+1):-said(A, B, M, T)
```

Next we define with the help of a choice rule the receiving of messages by an agent the protocol. The rule below suggests that if A sends the message msg(M) to B at time T, then B may/may not receive it. This relieves us from explicitly modeling the faulty transmission behavior or message interception.

```
gets(B, M, T):-says(A, B, M, T)
```

We use similar choice rules to describe the abilities of an intruder. In this protocol description and intruder spy can eavesdrop and receive any message from a protocol run. It can also transmit valid messages to honest agents, given that it is able to synthesize them.

```
gets(spy, M, T):-says(A, B, M, T), A!=spy, B!=spy
says(spy, B, M, T):- synth(spy, M, T), B!=spy
```

We then define the rules that specify a protocol's action. This being a protocol dependant part, we have to ensure that desirable constraints are imparted on our solution set. To do this we put the same preconditions to each action in a protocol run, as assumed in our initial description of the protocol. We also specify message validation rules, which enable the agents to proceed in a protocol run, only if a message or its component has been verified.

```
says(A,B,encrypt(K, Na, A),T):-
        fresh(Na, T), publicKey(K, B), A!=B

says(B,A,encrypt(K, Na, Na,T):-
      got(B,encrypt(K, Na, A),T), fresh(Nb, T),
      publicKey(K, A), A!=B

says(A,B,encrypt(K, Nb),T):-
      said(A,B,encrypt(K, Na, A),T),
      got(A,encrypt(K, Na, Na,T),
      publicKey(K, B), A!=B
```

This completes our background and action theories for the ALSP specification of the NSPK protocol. In the coming section we formally verify the protocol with the above specification and validate security claims earlier proposed by the protocol.

**Planning Attacks on the NSPK**

We specify goals using logic programming *rules* and execute our specification to see if there exists a model where the particular goal state is attained. If true, we deduce that the protocol can be manipulated to attain the security violation and state declaratively that the protocol is insecure. Similarly if a model is not generated we claim that the security violation can not be achieved hence, the protocol is secure. Again, we incorporate the approach as suggested in [6, 10] to specify our goals in ALSP. For example, a logic programming rule 'attack1' representing a state when an adversary has attained a sessions key by manipulating a protocol run could be written as:

```
attack1(T):-
      got(B,encrypt(Kb,Na),T),
      said(B,A,encrypt(Ka,Na,Nb),T),
      got(B,encrypt(Kb,Nb),T),
      not said(A,B, encrypt(Kb,Na),T),
      not  said(A,B,  encrypt(Kb,Nb),T),   A!=spy,
      B!=spy.
```
Result Set:
```
C:>lparse nspk.lp attack1.lp | smodels
smodels version 2.26. Reading...done
Answer: 1
True
Duration 109.827
Number of choice points: 208
Number of wrong choices: 24
Number of atoms: 125348
Number of rules: 1040351
Number of picked atoms: 254233
Number of forced atoms: 436
Number of truth assignments: 40149094
Size of searchspace (removed): 752 (221)
```

The model generated corresponds to that suggested in section III for Lowe's [4] attack on NSPK.

*C. THE SEQUENTIAL PROGRAMMING APPROACH*

CSP (Communicating Sequential Process) is a process algebra notation, for analysis of interaction between two or more processes, or between a process and its external environment [15, 16]. It allows a system to be described at any level of abstraction. FDR (Failure Divergence Refinement) [27] is a model checking tool based upon CSP theory for concurrent processes. FDR takes the system specification and implementation of a protocol in CSP script as input and produces a counterexample as output, if the implementation doesn't meet the given specification. FDR uses the technique of searching the state space to find any insecure sequences of messages that can occur leading to an attack on the protocol. Thus, FDR can only be used for finite systems. Gavin Lowe[4] used CSP and FDR to break and fix the Needham-Schroeder authentication protocol.

In order to model a protocol in CSP, each entity participating in the protocol is represented as a CSP process which communicate over channels. An intruder that can interact with the protocol is also modeled as a CSP process. Conventionally, the channel's names are of the form x(in/out)y, where in/out refers to the direction (relative to x), and y is the other party of the communication. For details on the most frequently used notations, please refer Table 1. For the complete list, refer [15, 16].

**Analysis of Needham Schroeder Protocol using CSP and FDR:**
An agent can either act as initiator (Send) or responder (Resp) of the protocol.

```
User(id,ns) = if ns == <> then STOP else
              Send(id,ns) [] Resp(id,ns)
```

The initiator agent chooses the agent with whom it wants to establish a session and then communicates the three messages of the NSPK protocol with it. After the three messages, the initiator enters a state wherein a session is established between the two parties. Similarly, the responder performs the three messages.

```
Send(id,ns) = |~| a:diff(agents,{id}) @
comm.id.a.pke(pk(a),Sq.<head(ns),id>) ->
([] n:nonces @

comm.a.id.pke(pk(id),Sq.<n,head(ns)>) ->
                  comm.id.a.pke(pk(a),n) ->
                  Session(id,a,n,tail(ns)))

Resp(id,ns) = [] a:diff(agents,{id}) @
            [] n:nonces @
              comm.a.id.pke(pk(id),Sq.<n,a>) ->

comm.id.a.pke(pk(a),Sq.<head(ns),n>) ->
                  comm.a.id.pke(pk(id),head(ns))
->Session(id,a,head(ns),tail(ns))
```

The set of messages are specified as follows. Here message4 is included to check for secrecy of nonces.

```
message1 = {pke(k,Sq.<n,a>) | k <- publickey, n <-
nonces, a <- agents}
comm1 = {a.b.m |  m <- message1, a<-agents, b<-
agents, a!=b}
message2 = {pke(k,Sq.<n,n'>) | k <- publickey, n <-
nonces, n' <- nonces}
comm2 = {a.b.m |  m <- message2, a<-agents, b<-
agents, a!=b}
message3 = {pke(k,n) | k <- publickey, n <- nonces}
comm3 = {a.b.m |  m <- message3, a<-agents, b<-
agents, a!=b}
message4 = {encrypt(n,m) | n <- nonces, m <-
wholemess}
comm4 = {a.b.m |  m <- message4, a<-agents, b<-
agents, a!=b}
```

An intruder can listen to messages between Alice and Bob, can interact with them, and can even intercept and fake messages. An intruder can also deduce facts to build messages. A deduction is a pair (X, a) where X is a finite set of facts and a is a fact which can be constructed using X. The three deductions rules – for sequencing, symmetric keys, and asymmetric keys – are specified as:

```
deductions1(X) = {({Sq . m}, nth(j,m)) ,
({nth(i,m) | i <-{0..#m-1}}, Sq . m) |
Sq.m <- X, j<-{0..#m-1}}
```

| Terms | Notation | Description |
|---|---|---|
| Process and Event | `P = a->Q` | process `P` performs event `a` and then behaves like process `Q` |
| External/Deterministic Choice | `a->P [] b->Q` | a process which can either perform event `a` and then behave like `P`, or perform event `b` and then behave like `Q`, according to whichever event (`a` or `b`) is first recorded |
| Internal/Nondeterministic Choice | `a->P |~| b->Q` | Same as external choice except the decision of choice is internal to the process and ambiguous to the environment |
| Input | `c ? x` | inputs value `x` from channel `c` |
| Output | `c ! x` | outputs value `x` on channel `c` |
| Concurrency | `P [|X|] Q` | processes `P` and `Q` synchronize on all events in `X` |
| Interleaving | `P [|{}|] Q` | processes `P` and `Q` run completely independent of each other |
| Indexed External Choice | `[]x:Z @ x->P` | Equivalent to `a->P [] b->P [] c->P` for `Z = {a, b, c}` (similarly Indexed Internal Choice is defined) |
| Communication | `comm.v` | value `v` of the message is communicated on channel `comm` |
| Special Processes | `STOP` | does nothing, represents a deadlock |
| | `SKIP` | represents successful termination |

TABLE 4
TERMS AND NOTATIONS USED IN CSP SPECIFICATION OF NSPK

```
deductions2(X) = {({m, k}, encrypt(k,m) ) ,
({encrypt(k,m), k}, m) |
Encrypt.(k,m) <- X}

deductions3(X) = {({m, k}, pke(k,m) ) ,
({pke(k,m), dual(k)}, m) |
PK.(k,m) <- X}
```

The protocol goals – after a successful run of the protocol, the intruder should not possess secret nonces and a session is established between the communicating parties – are also specified as part of the CSP file. This file is given as input to the FDR which shows the Lowe's attack. For the complete code, please refer [16] and [17].

**Casper:**

Modeling in CSP has been proven to be tedious and error prone. Producing a CSP description of a protocol is very time-consuming, and also demands expertise in CSP. Casper [26], a modeling tool, was built by Lowe to generate the CSP code of a protocol from a more abstract description of it. The auto-generated CSP code by Casper can then be used for checking using FDR. Casper input file consists of two major parts – a generic definition of the way in which the protocol operates and a definition of the actual system to be checked. Each part contains several subsections like Free Variables, Intruder Information, Protocol Description, Specification, System, etc. with the line beginning with '#'.

In Casper, the sequence of messages for the Needham Schroeder protocol is defined as below. Message 0 says the environment sends the identity of B to A, implying A is the initiator of the protocol with B as responder.

```
#Protocol description
0.      -> A : B
1. A -> B : {na, A}{PK(B)}
2. B -> A : {na, nb}{PK(A)}
3. A -> B : {nb}{PK(B)}
```

The goal to be achieved by the protocol is specified as:
```
#Specification
Secret(A, na, [B])
Secret(B, nb, [A])
Agreement(A,B,[na,nb])
Agreement(B,A,[na,nb])
```
Please refer [17] for the complete code.

### D. THE BAN LOGIC APPROACH

BAN logic was proposed by Mike Burrows, Martín Abadí, Roger Needham in [22]. It allows the assumptions and goals of a protocol to be stated abstractly in belief logic. According to it messages send by any user contains his beliefs. It defines the rules which state which govern how the belief state is updated on receiving any message. For a successful run of the protocol the belief state of communicating parties should contain the protocol goals.

TABLE 5
NOTATION IN BAN LOGIC FOR NSPK PROOF

| $P \stackrel{Kab}{\leftrightarrow} Q$ | K is a good key for communication between participants P and Q. |
|---|---|
| $\#(Np)$ | Nonce value Np is fresh and hence valid. |
| $P \mid \sim X$ | P once said X. |
| $P \mid \equiv X$ | P believes in X. |
| $P \mid \Rightarrow X$ | P has jurisdiction over X. |
| $P \triangleleft X$ | P sees X. |
| $\dfrac{X}{Y}$ | If a participant believes in X then he/she believes in Y too. |

BASIC RULES

- Message Meaning Rule

$$\dfrac{P \triangleleft \{X\}_K, \; P \mid \equiv P \stackrel{K}{\leftrightarrow} Q}{P \mid \equiv Q \mid \sim X}$$

- Nonce Verification Rule

$$\dfrac{P \mid \equiv Q \mid \sim X, \; P \mid \equiv \#(X)}{P \mid \equiv Q \mid \equiv X}$$

- Jurisdiction Rule

$$\dfrac{P \mid \equiv Q \mid \Rightarrow X, \; P \mid \equiv Q \mid \equiv X}{P \mid \equiv X}$$

**Needham-Schroeder Protocol Analysis:** The original NSPK protocol without the idealisation has been discussed in section III. Corresponding idealised protocol from message 2 is as follows:

PROTOCOL AIM:

The aim of NSPK protocol is that the participants believe that they share a common secret Kab and also each participant should believe that the other participant also believe the same.

IDEALISED PROTOCOL:
2. $S \rightarrow A$: $\{Na, (A \stackrel{Kab}{\leftrightarrow} B), \#(A \stackrel{Kab}{\leftrightarrow} B), \{A \stackrel{Kab}{\leftrightarrow} B\}Kbs\}Kas$
3. $A \rightarrow B$: $\{A \stackrel{Kab}{\leftrightarrow} B\}Kbs$
4. $B \rightarrow A$: $\{Nb, (A \stackrel{Kab}{\leftrightarrow} B)\}Kab$ from B
5. $A \rightarrow B$: $\{Nb, (A \stackrel{Kab}{\leftrightarrow} B)\}Kab$ from A

ASSUMPTIONS:
1. $A \models A \stackrel{Kas}{\leftrightarrow} S$; $B \models B \stackrel{Kbs}{\leftrightarrow} S$
2. $S \models A \stackrel{Kas}{\leftrightarrow} S$; $S \models B \stackrel{Kbs}{\leftrightarrow} S$; $S \models A \stackrel{Kab}{\leftrightarrow} B$
3. $A \models (S \models\Rightarrow A \stackrel{Kab}{\leftrightarrow} B)$
4. $B \models (S \models\Rightarrow A \stackrel{Kab}{\leftrightarrow} B)$
5. $A \models (S \models\Rightarrow \#(A \stackrel{Kab}{\leftrightarrow} B))$
6. $A \models \#(Na)$; $B \models \#(Nb)$
7. $S \models \#(A \stackrel{Kab}{\leftrightarrow} B)$;
8. $B \models \#(A \stackrel{Kab}{\leftrightarrow} B)$

PROOF:
We will now use the assumptions and logical postulates and apply them to each message. With the help of message (2) and assumption (1) we can derive that.

$A \triangleleft \{Na, (A \stackrel{Kab}{\leftrightarrow} B), \#(A \stackrel{Kab}{\leftrightarrow} B), \{A \stackrel{Kab}{\leftrightarrow} B\}Kbs\}Kas$ – (1)

Using this and applying the logical postulate – message meaning rule we can say:

$A \models S \mid\sim \{Na, (A \stackrel{Kab}{\leftrightarrow} B), \#(A \stackrel{Kab}{\leftrightarrow} B), \{A \stackrel{Kab}{\leftrightarrow} B\}Kbs\}$ – (2)

Now we use assumption (6) and apply freshness rule:

$A \models \# \{Na, (A \stackrel{Kab}{\leftrightarrow} B), \#(A \stackrel{Kab}{\leftrightarrow} B), \{A \stackrel{Kab}{\leftrightarrow} B\}Kbs\}Kas$ – (3)

Using (1), (2) and (3) and applying nonce verification rule we get:

$A \models S \models (A \stackrel{Kab}{\leftrightarrow} B)$ and $A \models S \models \#(A \stackrel{Kab}{\leftrightarrow} B)$

Using assumptions (3, 5) we apply the jurisdiction rule to finally get:

**$A \models (A \stackrel{Kab}{\leftrightarrow} B)$ and $A \models \#(A \stackrel{Kab}{\leftrightarrow} B)$ - R1**

Now we move on to message (3) and apply logical postulates with the help of our assumptions.

$B \triangleleft \{A \stackrel{Kab}{\leftrightarrow} B\}Kbs$

Applying the message meaning rule we can derive:

$B \models S \mid\sim \{A \stackrel{Kab}{\leftrightarrow} B\}Kbs$

Using assumption (8) we apply nonce verification rule and get:

$B \models S \models \{A \stackrel{Kab}{\leftrightarrow} B\}$

Using assumption(4) and applying jurisdiction rule we get:

**$B \models \{A \stackrel{Kab}{\leftrightarrow} B\}$ - R2**

For message (4) we use the previously derived result (R1) and apply message meaning rule to get:

$A \models B \mid\sim \{(A \stackrel{Kab}{\leftrightarrow} B)\}$ – (4)

Again using (R1) and (4) we apply nonce verification rule to get our protocol aim.

**$A \models B \models (A \stackrel{Kab}{\leftrightarrow} B)$ and $B \models A \models (A \stackrel{Kab}{\leftrightarrow} B)$ - R3**

This protocol has an extra assumption, which is that B assumes the key B receives from A is fresh (assumption 8). So Needham-Schroeder protocol had this flaw in it.

In this context although B sends Nb encrypted with Ka, the assumption is not valid because any adversary is able to trick A into decrypting B's message. This weakness was exploited by Lowe in his attack.

$B \models \#(A \stackrel{Kab}{\leftrightarrow} B)$

V. QUALITATIVE COMPARISON OF FORMAL APPROACHES

The most important feature of approaches under the logical inference domain is its independence from *automated support*. They rely primarily on the assumptions taken before the analysis of a protocol and capture the notion of validity of data items such as *nonce* and *session keys*. Strand Spaces captures this notion with the help of two partial orderings, namely *subterm* and *preceq* [24]. The ease of usage of Strand Spaces over the BAN logic makes the former more user-friendly.

TABLE 6
COMPARING THE TWO LOGICAL INFERENCE APPROACHES

| BAN LOGIC | STRAND SPACES |
|---|---|
| BAN logic does not allow modeling of capabilities of system penetrator explicitly. | Strand Spaces model captures all the possible adversary behaviour independent of the protocol being analyzed. Thus all the capabilities and limitations of the adversary is taken under consideration beforehand |
| BAN logic analyses the protocol, one message at a time. | Strand Spaces analyses all the messages received by a party together, using *strands*. |
| BAN logic is not very reliable for the verification of security protocols, since it proved insecure protocols like NSPK and Ottway-Rees protocol [28] secure. | Strand Spaces on the other hand is able to detect flaws in the mentioned protocols. |

Both the logic programming and the sequential programming approaches are capable of verifying the NSPK protocol for Lowe's attack. An advantage of using ALSP and CSP over inference logic approaches is that these specification languages allow a system to be described at any level of abstraction. It is also worthwhile to state that, the development of a protocol specification in ALSP is notably easier than in CSP. The latter incorporates far typical syntaxes as compared to the former.

TABLE 7
COMPARING THE TWO MODEL CHECKING APPROACHES

| CSP/FDR | ALSP |
|---|---|
| FDR used with Casper allows generating the attacking run corresponding to prospective security flaws. | ALSP specification when executed on the *smodels* model finder generates a protocol trace corresponding to security violations. Often, the generation of an attacking run from the trace itself proves to be a tedious task. |
| Complex security properties like *atomicity & fairness* can easily be specified using sequential programming. | Specification of such properties using the logic programming approach is yet to be explored. |
| Casper enables a protocol designer to specify the protocol in the prevalent Alice-Bob notation. This considerably reduces the time and efforts employed in the protocol verification. | Tools to generate the ALSP code of a protocol from a more abstract description of it is not yet been designed. Thus, the development of a specification in ALSP is more time consuming. |
| FDR cannot deal with infinite state systems. Thus, CSP cannot check any arbitrary protocol. | The semantics for protocol specification in ALSP ensures that the specification is admissible. Therefore, a case of an infinite state system is out rightly rejected. |

## VI. FUTURE WORK

Our intention of future work includes identifying common parameters like computational costs, for comparison of different formal approaches for verification. We would also extend our work by qualitatively comparing other approaches like GNY Logic, Scyther, Murphy and AVISPA.

## VII. CONCLUSION

We are able to identify and compare the basic limitations and advantages of using various formal verification approaches for security protocol analysis. This enables us to choose a suitable verification approach catering to a particular given scenario. For example, in absence of any automated support, we can use a logical inference over other model checking approaches. For infinite state systems, ALSP is preferred over CSP. As far as ease of usage is concerned Casper scores over others. BAN logic is not a preferred approach as it has been proven to give false result for few protocols. To capture complex goals required for fair exchange protocols, CSP and Strand Spaces offer a suitable solution.